\documentclass[twocolumn,showpacs,preprintnumbers,superscriptaddress,amsmath,amssymb,floatfix,prl]{revtex4}

\usepackage{graphicx}
\usepackage{dcolumn}
\usepackage{bm}
\usepackage{soul}

\usepackage[utf8]{inputenc}
\usepackage{xcolor}

\usepackage{float}

\newcommand{\vkO}{\overrightarrow{k}_0}
\newcommand{\vkp}{\overrightarrow{k}_p}
\newcommand{\vkSRS}{\overrightarrow{k}_{SRS}}

\begin{document}


\title{\textbf{Electromagnetic emission at the Langmuir frequency\\from laser interaction with an unmagnetized plasma}}

\author{J.-R. Marquès}
\email[]{jean-raphael.marques@polytechnique.fr}
\affiliation{LULI, CNRS, CEA, Sorbonne Université, École Polytechnique, Institut Polytechnique de Paris, 91128 Palaiseau, France}

\author{F. Pérez}
\affiliation{LULI, CNRS, CEA, Sorbonne Université, École Polytechnique, Institut Polytechnique de Paris, 91128 Palaiseau, France}

\author{P. Loiseau}
\affiliation{CEA, DAM, DIF, F-91297 Arpajon, France}
\affiliation{Université Paris-Saclay, CEA, LMCE, 91680 Bruyères-le-Châtel, France}

\author{L.~Lancia}
\affiliation{LULI, CNRS, CEA, Sorbonne Université, École Polytechnique, Institut Polytechnique de Paris, 91128 Palaiseau, France}

\author{C. Briand}
\affiliation{LIRA, Observatoire de Paris-PSL, CNRS, Sorbonne Université, Université de Paris, 92195 Meudon, France}

\author{S. Depierreux}
\affiliation{CEA, DAM, DIF, F-91297 Arpajon, France}

\author{M. Grech}
\affiliation{LULI, CNRS, CEA, Sorbonne Université, École Polytechnique, Institut Polytechnique de Paris, 91128 Palaiseau, France}

\author{C. Riconda}
\affiliation{Sorbonne Université, LULI, CNRS, CEA, École Polytechnique, Institut Polytechnique de Paris, 75255 Paris, France}

\date{\today}

\begin{abstract}
We present the first experimental observation of electromagnetic emission at the Langmuir frequency, $\omega_p$, resulting from laser-plasma interaction in an underdense and unmagnetised plasma. The emission is spatially resolved and analyzed in polarization. It originates from an electrostatic  wave excited by the Raman instability. Despite its frequency being very close to the plasma's cutoff frequency, $\omega_{pe}$, this radiation can escape the plasma by following the decreasing slope of the plasma density profile. Radiation at $2\omega_p$, is also observed. This experimental platform offers a powerful tool to explore the processes of wave coupling at the origin of some radio solar bursts.

\end{abstract}


\maketitle

Measuring electromagnetic (EM) radiation at a frequency $\omega_p$ close to the electron plasma frequency $\omega_{pe}$ is a powerful diagnostic tool in many areas of plasma physics. In astrophysics, solar radio bursts known as Type II and Type III bursts have been observed in interplanetary space for decades \cite{Reid,Pick,Cairns2003}. These bursts consist of radio waves (10–300 MHz) at the local plasma frequency or its first harmonic ($2\omega_p$). They originate from Langmuir waves (plasma electron oscillations) that are produced by beams of energetic electrons accelerated during solar eruptions. Through decay and coupling processes, the electrostatic (ES) Langmuir waves give rise to transverse EM waves. These waves carry essential information about the mechanisms underlying particle acceleration and energy release during solar flares and coronal mass ejections. They provide a powerful means of studying the solar corona and the interplanetary medium.

During the 1970s and 1980s, laboratory astrophysics experiments were conducted to investigate the mechanisms underlying solar radio bursts. EM radiation at $\omega_p$, and occasionally at $2\omega_p$, in the microwave range (0.3–250 GHz), were produced from theta pinch \cite{Chin-Fatt}, plasma discharges \cite{Means,Whelan1981, Whelan1985}, with in some cases the injection of a relativistic electron beam \cite{Kato, Benford, Cheung}. However, the limitations of these devices prevented more in-depth studies from being conducted. More recently (2014), in the quest for powerful THz sources, EM radiations at $\omega_p$ and $2\omega_p$ were generated in the 0.1-0.4 THz range by injecting a relativistic electron beam into a magnetized plasma \cite{Arzhannikov}.

In a magnetized plasma, a Langmuir wave has both ES and EM components (Langmuir$/$Z-mode wave). This enables the wave to couple radiation at the plasma-vacuum interface\cite{Yoshii, Yip}. However, the ES-to-EM conversion can also be produced without B-field, by linear mode conversion in a density gradient \cite{Means, Arzhannikov, Hinkel}, by the so-called antenna mechanism, where Langmuir waves are localized as eigenmodes of density cavities \cite{Malaspinal}, or by nonlinear three-wave processes. In this latter case, a Langmuir wave $L$ can decay in two ways, through the EM decay \cite{Cairns1987, Edney,Shukla}, $L \rightarrow S + F$, or through the ES decay \cite{Robinson,Palastro}, $L \rightarrow L' + S$. Here $F$ is a transverse EM wave at the fundamental frequency $\omega_p$, $S$ is an ion acoustic wave, and $L'$ is a daughter Langmuir wave. The ES decay can lead, in a second step, to radiation at the fundamental frequency and its first harmonic $H$ via the coupling processes $L + S \rightarrow F$ and $L +  L' \rightarrow H$, respectively. Depending on the plasma conditions, all of these mechanisms can coexist \cite{Krafft}.

The growth, decay and coupling of the plasma waves depend on plasma parameters such as the electron density $n_e$, temperatures $T_e$, the ion temperature $T_i$, charge $Z$ and mass $m_i$. The dominant conversion mechanism thus depends on these parameters, defining the properties of the $\omega_p$ radiation (power, angular distribution, polarization). In particular, the nonlinear three-wave processes depend heavily on the $S$ wave and thus, on the ion type. To our knowledge, over the last 56 years, only seven laboratory experiments on solar bursts have investigated radiation at $\omega_p$, most of which used heavy ions (Ar, Si, Ti). Only two studies \cite{Benford,Whelan1981} have been conducted in a hydrogen plasma, which is relevant to the solar corona and the interplanetary medium. Using plasmas of H, He, Ar and Xe, Whelan \textit{et al.} (1981) \cite{Whelan1981,Whelan1985} conducted the only experiment to verify the role of ions (via the $S$ wave) in $\omega_p$ radiation. This led them to claim that the $L + S \rightarrow F$ process was responsible for the $\omega_p$ radiation in their experimental conditions. However, another interpretation has been proposed, implying EM decay instead \cite{Shukla}. Furthermore, although they observed radiation at $2\omega_p$, its intensity was negligible despite solar bursts with similar intensities at $\omega_p$ and $2\omega_p$ often being observed.

Two experiments on laser-plasma interaction observed EM radiation at $\omega_p$, both using sub-ps lasers. EM radiations at $\omega_p$ and $2\omega_p$ in the VUV region were produced by focusing the laser onto an unmagnetized solid target (glass) \cite{Teubner_1997,Teubner_1999}. However, the high collisionality of overdense plasmas makes them unsuitable for studying solar bursts. In the context of new THz sources. Yugami \textit{et al.} \cite{Yugami} produced a near-THz (0.2 GHz) pulse from an underdense Ar plasma. The $\omega_p$ emission originated from a Cerenkov wake propagating in a perpendicular magnetic field \cite{Yoshii}, while solar bursts are mainly produced along the magnetic lines, by nonlinear three-wave processes.

Recently, we proposed a new experimental platform \cite{Marques2020} to explore the fundamental processes of wave coupling at the origin of solar bursts: rather than being excited by high-energy electrons via the bump-on-tail instability, the primary Langmuir wave $L$ is excited by the Raman instability of a nanosecond laser focused into an underdense gas jet. EM radiation at the harmonic frequency $2\omega_p$ from a hydrogen plasma was observed. The intensity, angular distribution, polarization and spectral evolution of this radiation confirmed the two-step scenario: $L \rightarrow L' + S$ followed by $L +  L' \rightarrow H$.

In this Letter we present the first observation of EM radiation at the fundamental plasma frequency, $\omega_p$, generated by a laser in an unmagnetized underdense plasma, via the Raman instability. The unexplored plasma density range we employed ($10^{20} < n_e < 10^{21}$ cm$^{-3}$) results in an $\omega_p$ emission in the near infrared (1.2-1.8 $\mu$m). This range enables conditions that match those of space plasmas \cite{Robinson} in terms of the dimensionless parameters \cite{Marques2020,Perez} governing nonlinear three-wave processes and the properties of the $F$ and $H$ radiations. Our experimental results are supported by particle-in-cell simulations (PIC), demonstrating that $F$ and $H$ radiations are produced, and have similar intensities. Hydrodynamic simulations, coupled with ray tracing, show that despite its frequency being very close to the plasma cutoff frequency $\omega_{pe}$, the $F$ radiation can escape the plasma by following a negative density gradient, with transmission of up to 75 $\%$.

The experiment was performed at the LULI2000 laser facility (École Polytechnique, France). The setup is shown in figure \ref{Fig_Setup} in the End Matter. A laser beam with a wavelength of $\lambda_0$ = 526.5 nm is focused at the center of a 0.5 mm diameter supersonic hydrogen gas jet. The laser is focused by a lens coupled to a hybrid phase plate, resulting in a focal spot with a speckle pattern within a hyper-Gaussian envelope of $\sim$ 500 $\mu$m diameter. The average intensity in the central region is $\sim 1.6\times 10^{14}$ W/cm$^2$, while the maximum speckle intensity is $\sim 8.5\times 10^{14}$ W/cm$^2$. The laser is linearly polarized at 45$^{\circ}$ with respect to the horizontal plane. An image of the interaction region is obtained by collecting the light at 85$^{\circ}$ onto a gated optical imager (GOI). The scattered light is also analyzed in time and frequency using spectrometers coupled to streak cameras. These are placed at angles of 25$^\circ$, 40$^\circ$, 75$^\circ$, 110$^\circ$ and 170$^\circ$ from the laser axis, in the horizontal plane. All these diagnostics cover the spectral range $700-900$ nm, where SRS and $2\omega_p$ emissions are expected. To measure the $\omega_p$ emission, the light collected at 75$^\circ$ and 110$^\circ$ is also recorded on an InGaAs camera to image the interaction region in the 1100-1700 nm spectral range. A Wollaston prism is placed in front of the camera to analyze the polarization of this infrared light. For each shot the camera therefore records four images, the horizontal and vertical polarization components of the light emitted at 75$^\circ$ and 110$^\circ$.

\begin{figure}
	\begin{center}
		\includegraphics[width=\columnwidth]{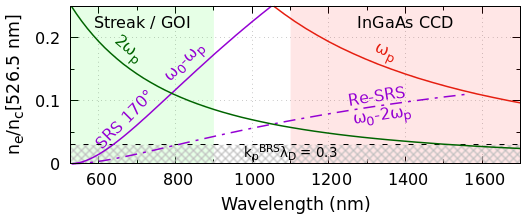}
		\caption{\small{\textit{Correspondence between the electron plasma density $n_e$ (in units of $n_c$) and the wavelengths of the EM waves from backward Raman scattering of the laser, Raman rescattering, and from emissions at $\omega_p$ and at $2\omega_p$. The green area shows the detection domain of the streak and GOI cameras, the red area of the InGaAs camera. The gray dashed area shows the density range in which the Langmuir wave produced by backward Raman scattering undergoes Landau damping.}}}
		\label{Fig_Wavelenghts_vs_ne}
	\end{center}
\end{figure}

Stimulated Raman scattering satisfies the conditions of momentum conservation, $\vkO = \vkp + \vkSRS$ and the energy conservation, $\omega_0 = \omega_p + \omega_{SRS}$. The subscripts $0$, $p$ and $SRS$ correspond to the incident laser, the Langmuir wave and the Raman scattered EM wave respectively. The Langmuir wave satisfies the dispersion relation $\omega_p = \omega_{pe}(1+3k_p^2\lambda_D^2)^{1/2}$, where $\lambda_D$ is the Debye length. Since an EM wave of frequency $\omega$ can propagate in the plasma only if $\omega > \omega_{pe}$, SRS can exist and grow only for $\omega_{pe} < \omega_0/2$, that is for $n_e < n_c/4$, where $n_c$ is the critical density beyond which the plasma becomes opaque to the laser ($\omega_{pe} > \omega_0$). On the other side, at low density, SRS is inhibited by Landau damping when $k_p\lambda_D \gtrsim 0.3$ \cite{LaFontaine}. In our experiment, the plasma expands over time, resulting in a decrease in density (decrease in $\omega_p$). The frequency of the SRS light, $\omega_{SRS} = \omega_0 - \omega_p$, and that of the EM emissions at $\omega_p$ and $2\omega_p$ thus evolve in time, with opposite behaviors. This is depicted in  Fig. \ref{Fig_Wavelenghts_vs_ne}: the wavelength of the SRS light decreases with density (time) while the wavelengths of the $\omega_p$ and $2\omega_p$ emissions increase.
Using our wavelength- and time-resolved diagnostics (green area in Fig. \ref{Fig_Wavelenghts_vs_ne}) the SRS and $2\omega_p$ emissions can thus be identified by the evolution of their wavelengths. The $\omega_p$ radiation is in the short-wave infrared domain (SWIR, red area in Fig. \ref{Fig_Wavelenghts_vs_ne}), in which no commercial streak camera is currently available. Its detection was performed using an InGaAs camera, which integrates the signal over a $\mu$s timescale. Unfortunately, due to the plasma expansion, when $n_e/n_c < 0.1$, both the Raman rescattering (Re-SRS, $\omega = \omega_0 - 2 \omega_p$) \cite{Marques2025} and the $2\omega_p$ radiations can enter the camera's detection window, preventing a proper isolation of the $\omega_p$ signal. This critical issue was resolved by limiting the density range of the interaction via the laser duration.

\begin{figure}
	\begin{center}
		\includegraphics[width=\columnwidth]{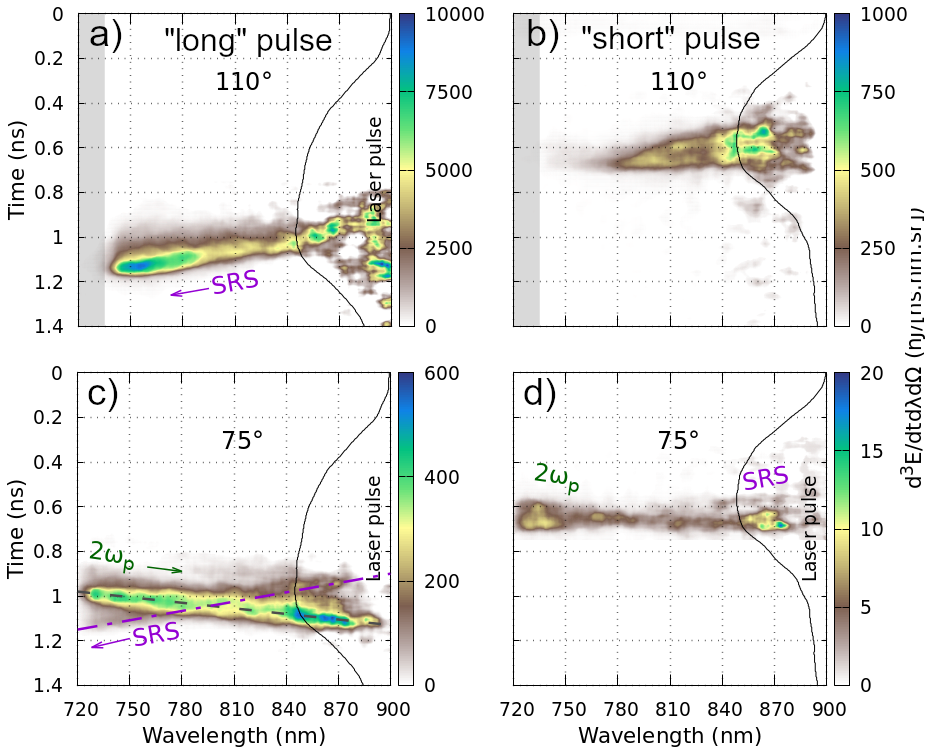}
		\caption{\small{\textit{Time-resolved spectra measured at 110$^\circ$ and 75$^\circ$ from the laser axis, for two similar shots, one with a "long" laser pulse, a) and c), and one with a "short" pulse, b) and d). The black curve on the right of each graph shows the temporal profile of the laser. The backing pressure is 600 bars, leading to an initial electron plasma density of $\sim 2.4\times 10^{21}$ cm$^{-3}$ ($n_e(t=0) \sim 0.6 n_c$). Each spectrum is corrected from the spectral response of the diagnostic. In a) and b) the cut below $\sim$ 730 nm (gray area) comes from a instrumental cutoff.}}}
		\label{Fig_Spectra}
	\end{center}
\end{figure}

Figure \ref{Fig_Spectra} shows the time-resolved frequency spectra obtained for two laser pulse durations (one "long" and one "short") and two angles relative to the laser axis (75$^\circ$ and 110$^\circ$). The gas jet is fully ionized at the very beginning of the laser pulse. The plasma is then heated via inverse bremsstrahlung and begins to expand. During this expansion, the plasma frequency decreases, causing the SRS radiation to shift towards shorter wavelengths, as illustrated in Figure \ref{Fig_Wavelenghts_vs_ne}. This phenomenon is clearly visible in the spectra at 110$^\circ$ (Fig. \ref{Fig_Spectra}-a and \ref{Fig_Spectra}-b). At 75$^\circ$ (Fig. \ref{Fig_Spectra}-c and \ref{Fig_Spectra}-d), the SRS signal is weaker, enabling to also observe the  radiation at $2\omega_p$, whose wavelength increases during plasma expansion. Comparing the left and right plots shows that SRS and $2\omega_p$ emissions cease at the falling edge of the laser pulse, indicating that Langmuir waves do not persist beyond the pulse. In the case of the "short" pulse, Fig. \ref{Fig_Spectra}-b and \ref{Fig_Spectra}-d, the SRS emission stops at a wavelength near 780 nm, corresponding to a plasma density of $n_e/n_c \sim$ 0.1. The disappearance of Langmuir waves is also observed in the spectra at 170$^\circ$, see Fig. \ref{Fig_Spectra_BRS}-b) in the End Matter. This is to be expected since when the laser intensity falls, the Langmuir waves are no longer excited. At this density and for $T_e\sim$ 500 eV, the damping timescale of a Langmuir wave is a few ps. Therefore, limiting the duration of the laser pulse enables the excitation of Langmuir waves to be restricted to plasma densities between 0.25 and 0.1 $n_c$. This allows radiation at only $\omega_p$ to fall within the detection window of the InGaAs camera, as depicted in Fig. \ref{Fig_Wavelenghts_vs_ne}.

\begin{figure}
	\begin{center}
		\includegraphics[width=\columnwidth]{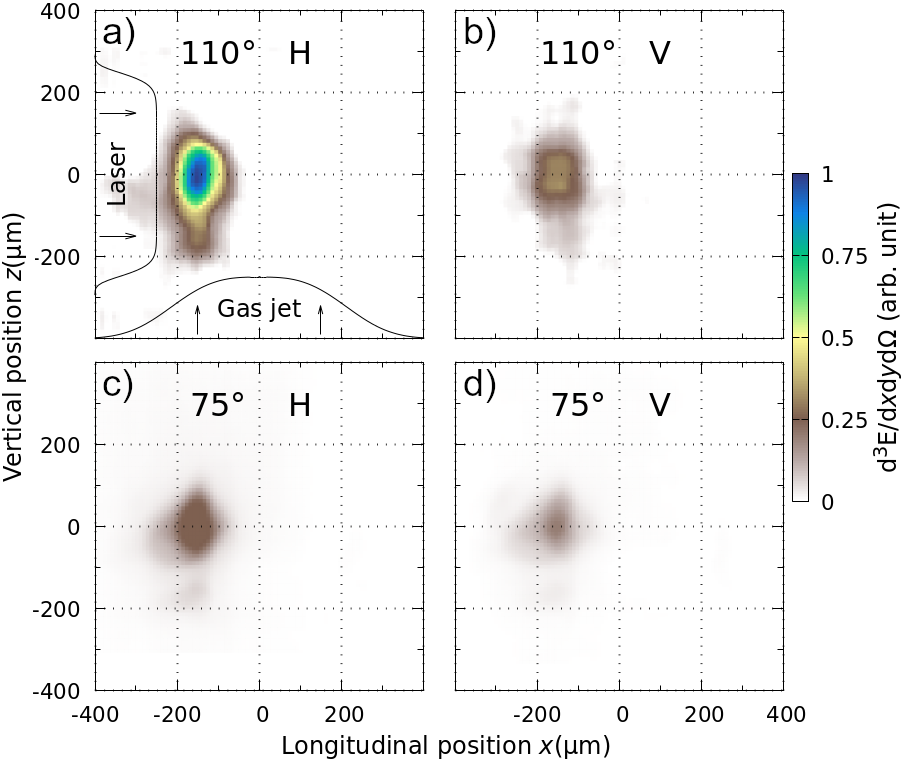}
		\caption{\small{\textit{2D-image (side view) of the interaction region in the "short" pulse case, obtained from the InGaAs camera (1100-1700 nm), from the scattered light emitted at 110$^\circ$ (top panel) and 75$^\circ$ (bottom panel) of the laser axis, for the horizontal (left panel) and vertical (right panel) components of the polarization. The laser propagates from left to right. The black curves show the transverse profile of the gas jet and of the laser focal spot (averaged on the speckles, see Fig. \ref{Fig_Setup}).}}}
		\label{Fig_InGaAs_image}
	\end{center}
\end{figure}

The signal obtained from the InGaAs cameras for the case of this "short" pulse (Fig. \ref{Fig_Spectra}-b) and \ref{Fig_Spectra}-d)) is presented in Fig. \ref{Fig_InGaAs_image}. As explained above, 4 images of the interaction region are recorded on the detector for each shot: the light emitted at 75$^\circ$ and 110$^\circ$, and for both their horizontal and vertical polarization components. The signal is centered on the laser axis and $\sim$ 150 $\mu$m upstream from the focal plane (gas jet center), at the same position as the SRS and $2\omega_p$ emissions observed in the visible-NIR domain with the GOI detector at 85$^{\circ}$, see Fig. \ref{Fig_Sideview} in the End Matter. The signal on the InGaAs camera is generated in the short laser pulse configuration, i.e. when $n_e/n_c > 0.11$. The wavelengths of the SRS, $2\omega_p$ and Re-SRS emissions are thus outside the camera's detection window (red area in Fig. \ref{Fig_Wavelenghts_vs_ne}). The signal is predominantly polarized horizontally, which confirms that it does not originate from SRS or Re-SRS, as these would follow the laser polarization, which is mainly vertical when observed at 90$^{\circ}$ from the laser axis. Furthermore, the spatial extent of the IR emission observed in Fig. \ref{Fig_InGaAs_image} is smaller than that associated with the SRS and the $2\omega_p$ signals measured on the GOI camera (Fig. \ref{Fig_Sideview} in the End Matter). Despite the camera's long integration time ($\mu$s), which covers the total plasma expansion, the emission is very localized and smaller than the laser focal spot and the gas jet. This indicates that the signal cannot originate from bremsstrahlung emission of the plasma. All these observations suggest that the observed signal originates from EM emission at $\omega_p$.

\begin{figure}
	\begin{center}
		\includegraphics[width=\columnwidth]{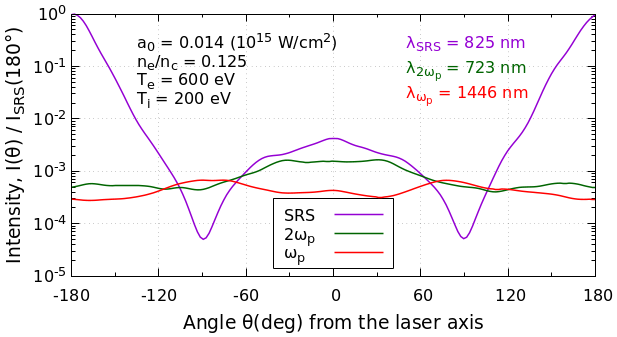}
		\caption{\small{\textit{Angular distribution of the SRS, $\omega_p$ and $2\omega_p$ signals obtained from Smilei PIC simulations. The intensities are normalized to the SRS maximum.}}}
		\label{Fig_Smilei}
	\end{center}
\end{figure}

To verify that, in our experimental conditions, EM emission at the Langmuir frequency is generated and can escape the plasma, we performed two-dimensional simulations: kinetic with the Smilei code \cite{Smilei} and hydrodynamic with the TROLL code \cite{TROLL}. Details of both simulations and additional results can be found in the Supplementary Material \cite{SM}. The measured experimental parameters (laser energy, focal spot and temporal profiles, and gas jet density profile) were used as inputs. The outputs of the hydrodynamic simulations were used to initialize the plasma parameters ($n_e$, $T_e$, $T_i$) for the PIC simulations. A previous study \cite{Perez} demonstrated that the growth and coupling of plasma waves are mainly governed by the most intense parts of the focal spot. Therefore, the PIC code simulates the interaction of a single speckle within the global focal spot (see Fig. \ref{Fig_Setup}). The speckle has a waist of 4.2 $\mu$m and a peak intensity of $10^{15}$ W/cm$^2$. The interaction is simulated for up to 10 ps, which is longer than the time required for the Raman and ES decay instabilities to grow and saturate \cite{Perez}. A result of these simulations is presented in Fig. \ref{Fig_Smilei} for plasma parameters associated with the "short" pulse case in Fig. \ref{Fig_Spectra}. It shows the angular intensity distributions of the EM emissions, normalized to the SRS intensity at 180$^{\circ}$. The density is $n_e/n_c \sim$ 0.125, at which the wavelengths of the SRS and the $2\omega_p$ radiations are close to those observed shown in Fig. \ref{Fig_Spectra}-d). The intensity ratios between the emissions are in good agreement with the experimental results shown in Fig. \ref{Fig_Spectra}: the SRS light dominates significantly in the backward direction, while at 75$^{\circ}$ the SRS and $2\omega_p$ radiations have similar intensities. This simulation also shows that the EM radiation at $\omega_p$ is indeed excited, with an intensity similar to that of the $2\omega_p$ radiation, and with an almost isotropic angular distribution. Simulations run at other densities ($n_e/n_c$ = 0.14, 0.09 and 0.05) lead to the same conclusions, and indicate that at $n_e/n_c < 0.1$, EM radiation from Re-SRS appears, but its intensity near 90$^{\circ}$ is at least two orders of magnitude lower than the $\omega_p$ and $2\omega_p$ radiations.

\begin{figure}[h!]
	\includegraphics[width=\columnwidth]{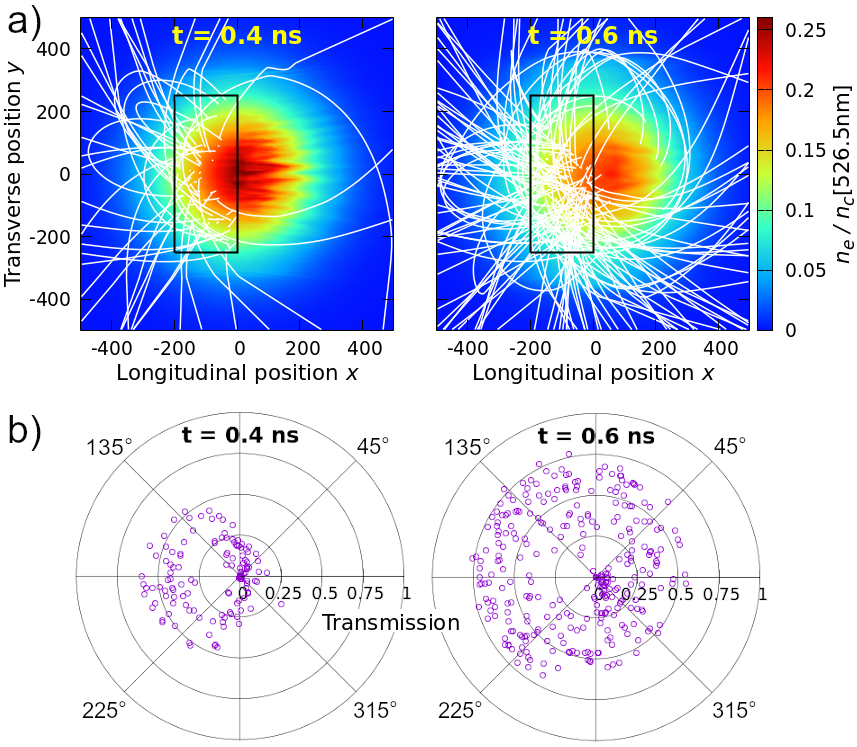}
	\caption{\small{\textit{a): 2D (top view) spatial profiles of the plasma density obtained from hydrodynamic simulations, at two instants in the laser pulse (black curve in Fig. \ref{Fig_Spectra} b) and d)), and trajectories of light rays (white lines) emitted at the local Langmuir frequency $\omega_p(x,y)$ from the region of interest (ROI) indicated by the rectangle, and in the detection window of the InGaAs camera (1100-1700 nm). The laser propagates from left to right. The ROI corresponds to the laser transverse size (along $y$) and, along the longitudinal position $x$, to the localization of the emitting zone observed in Fig. \ref{Fig_InGaAs_image}. Only the rays with a non-zero transmission are plotted. b): Transmission of all the rays versus the output angle. For $t<0.3$ ns and $t>1$ ns the wavelengths of the $\omega_p$ rays are outside the detection window.}}}
	\label{Fig_Ray_tracing}
\end{figure}

These PIC simulations confirm that the laser-plasma interaction produces EM radiation at the Langmuir frequency $\omega_p$. However, as this frequency is very close to the plasma cutoff frequency $\omega_{pe}$, the radiation can undergo significant absorption and refraction when propagating through the plasma. To evaluate these effects, we combined the results of the TROLL hydrodynamic simulations with a ray tracing post-processing. First, the two-dimensional spatial profiles of $n_e$, $T_e$ and $T_i$ are extracted. A set of $\omega_p$ rays is then generated and propagated through the density gradients. The initial position of each ray, ($x_i,y_i$), is randomly sampled from a region of interest representative of the regions where the $\omega_p$, the $2\omega_p$ and the SRS emissions occur, as observed in Fig. \ref{Fig_InGaAs_image} and in Fig. \ref{Fig_Sideview} in the End Matter. The initial direction is randomly sampled from an isotropic distribution (see PIC simulations, Fig. \ref{Fig_Smilei}). The frequency of each ray, $\omega_p = \omega_{pe}(1+3k_p^2\lambda_D^2)^{1/2}$, is determined by the values of $\omega_{pe}$ and $k_p \lambda_D$ at its initial position, which depend on $n_e(x_i,y_i)$ and $T_e(x_i,y_i)$. The wavenumber $k_p(x_i,y_i)$ is that of the Langmuir wave from backward SRS, which is at the origin of the EM emissions. Only rays with wavelengths within the detection window of the InGaAs camera are retained. These rays then undergo refraction and bremsstrahlung absorption. The results of these simulations are presented in Fig. \ref{Fig_Ray_tracing} at two times in the laser pulse. The $\omega_p$ rays can undergo strong refraction and absorption. The trajectories of those with non-zero transmission are plotted in Fig. \ref{Fig_Ray_tracing}-a). The transmission of all the rays versus their output angles is presented in Fig. \ref{Fig_Ray_tracing}-b). At $t$ = 0.4 ns the center of the plasma is still dense and exhibits some small-scale inhomogeneities produced by the filamentation of the laser (see SM \cite{SM}). Only the rays propagating in the backward half-plane exit the plasma. These rays follow the decreasing slope of the density profile, meaning their frequency is always greater than the local cutoff frequency. At $t$ = 0.6 ns, where in Fig. \ref{Fig_Spectra}-d) the SRS and $2\omega_p$ signals are observed, the plasma has started to expand. Its density is lower and the gradients are smoother, allowing more rays to exit the plasma, and with a higher transmission, up to 75 $\%$, even near 90$^{\circ}$.
These ray tracing simulations confirm that despite its frequency close to the plasma cutoff frequency the radiation at $\omega_p$ can escape the plasma.

In conclusion, we present the first experimental observation of electromagnetic (EM) emission at the Langmuir frequency, $\omega_p$, resulting from laser-plasma interaction in an underdense and unmagnetized plasma. This emission originates from an electron plasma wave that is excited by the Raman instability. Despite its frequency being very close to the plasma's cutoff frequency, $\omega_{pe}$, the radiation can escape the plasma with transmission of up to tens of percent, by following the decreasing slope of the plasma density profile. This is analogous to the propagation of solar bursts through the density gradient of the solar corona. EM emission at $2\omega_p$ is also observed, like in the astrophysical scenario. The range of plasma densities used here is suitable for upcoming experiments that will achieve a level of magnetization representative of space conditions $\omega_c /\omega_p\sim0.01$, where $\omega_c$ is the cyclotron frequency of the electrons, via magnetic fields of a few tens of Tesla. Such fields can be produced, for example, with pulsed coils \cite{Albertazzi}. This laser platform can therefore closely mimic the dimensionless parameters ($k_p\lambda_D$, $T_e/T_i$, $k_{EM}/k_p$, $\omega_c/\omega_p$) that govern solar bursts, enabling parametric studies to better understand the generation of the $\omega_p$ and $2\omega_p$ radiations, their respective intensities, angular distribution, polarization, as well as the influence of the magnetic field.

\begin{acknowledgments}
\textit{Acknowledgments} -- The authors thank the invaluable support of the LULI2000 staff, and the development team of the PIC code Smilei for their help. Smilei simulations were performed with computing and storage resources by GENCI at CINES thanks to the grant 2024-A0170507678 on the supercomputer Adastra's MI250x partition.
\end{acknowledgments}

\textit{Data availability} -- The data that support the findings of this Letter are not publicly available upon publication because it is not technically feasible and/or the cost of preparing, depositing, and hosting the data would be prohibitive within the terms of this research project. The data are available from the authors upon reasonable request.

\bibliography{bibliography}

@article{Reid,
    author = {Reid, H. A. S. and Ratcliffe, H.},
    title = {A review of solar type III radio bursts},
    journal = {RAA},
    volume = {14},
    pages = {773},
    year = {2014},
    doi = {10.1088/1674–4527/14/7/003},
    url = {https://doi.org/10.1088/1674-4527/14/7/003}
}

@article{Pick,
    author = {Pick, Monique and Vilmer, Nicole},
    title = {Sixty-five years of solar radioastronomy: flares, coronal mass ejections and Sun Earth connection},
    journal = {Astronomy \& Astrophysics Review},
    volume = {16},
    pages = {1-153},
    year = {2008},
    doi = {10.1007/s00159-008-0013-x},
    url = {}
}

@article{Cairns2003,
	author = {CAIRNS, IVER H. and KNOCK, S. A. and ROBINSON, P. A.  and KUNCIC, Z.},
	title = {TYPE II SOLAR RADIO BURSTS: THEORY AND SPACE WEATHER IMPLICATIONS},
	journal = {Space Science Reviews},
	volume = {107},
	pages = {27},
	year = {2003},
	doi = {},
	url = {https://doi.org/10.1023/A:1025503201687}
}

@article{Krafft,
	author = {Krafft, C. and Savoini, P. and Polanco-Rodríguez, F. J.},
	title = {Mechanisms of Fundamental Electromagnetic Wave Radiation in the Solar Wind},
	journal = {The Astrophysical Journal Letters},
	volume = {967},
	pages = {L20},
	year = {2024},
	doi = {},
	url = {https://doi.org/10.3847/2041-8213/ad47b5}
}

@article{Chin-Fatt,
    author = {Chin-Fatt, C. and Griem, H. R.},
    title = {Enhanced Radiation from a Theta-Pinch Plasma},
    journal = {Phys. Rev. Lett.},
    volume = {25},
    pages = {1644},
    year = {1970},
    doi = {},
    url = {https://doi.org/10.1103/PhysRevLett.25.1644}
}

@article{Means,
    author = {Means, R. W. and Muschietti, L. and Tran, M. Q. and Vaclavik, J.},
    title = {Electromagnetic radiation from an inhomogeneous plasma: Theory and experiment},
    journal = {Phys. Fluids},
    volume = {24},
    pages = {2197},
    year = {1981},
    doi = {10.1063/1.863326},
    url = {https://aip.scitation.org/toc/pfl/24/12}
}

@article{Whelan1981,
	author = {Whelan, D. A. and Stenzel, R. L.},
	title = {Electromagnetic-wave excitation in a large laboratory beam-plasma system},
    journal = {Phys. Rev. Lett.},
    volume = {47},
    pages = {95},
    year = {1981},
	doi = {10.1103/PhysRevLett.47.95},
	url = {http://dx.doi.org/10.1103/PhysRevLett.47.95}
}

@article{Whelan1985,
	author = {Whelan, D. A. and Stenzel, R. L.},
	title = {Electromagnetic radiation and nonlinear energy flow in an electron beam-plasma system},
    journal = {Phys. Fluids},
    volume = {28},
	pages = {958},
	year = {1985},
	doi = {10.1063/1.865067},
    url = {http://dx.doi.org/10.1063/1.865067}
}

@article{Shukla,
	author = {Shukla, P. K. and Yu, M. Y. and Mohan, M. and Varma, R. K. and Spatschek, K. H.},
	title = {Electromagnetic wave generation in a beam-plasma system},
	journal = {Phys. Rev. A},
	volume = {27},
	pages = {552},
	year = {1983},
	doi = {},
	url = {DOI: https://doi.org/10.1103/PhysRevA.27.552 }
}

@article{Benford,
    author = {Benford, G. and Tzach, D. and Kato, K. and Smith, D. F.},
    title = {Collective Microvrave Emission from Intense Elecron-Beam Interactions: Theory and Experiment},
    journal = {Phys. Rev. Lett.},
    volume = {45},
    pages = {1182},
    year = {1980},
    doi = {},
    url = {https://doi.org/10.1103/PhysRevLett.45.1182}
}

@article{Cheung,
	author = {Cheung, P. Y. and Wong, A. Y. and Darrow, C. B. and Qian, S. J.},
	title = {Simultaneous observation of caviton formation, spiky turbulence, and electromagnetic radiation},
	journal = {Phys. Rev. Lett.},
	pages = {1348},
	volume = {48},
	year = {1982},
	doi = {10.1103/PhysRevLett.48.1348}
}

@article{Kato,
    author = {Kato, K. G. and Benford, G. and Tzach, D.},
    title = {Detailed spectra of high‐power broadband microwave radiation from interactions of relativistic electron beams with weakly magnetized plasmas},
    journal = {Phys. Fluids},
    volume = {26},
    pages = {3636},
    year = {1983},
    doi = {},
    url = {https://doi.org/10.1063/1.864134}
}

@article{Arzhannikov,
	author = {Arzhannikov, A. V. and \textit{al.}},
	title = {Observation of spectral composition and polarization of sub-terahertz emission from dense plasma during relativistic electron beam–plasma interaction},
	journal = {Phys. Plasmas},
	volume = {21},
	pages = {082106},
	year = {2014},
	doi = {},
	url = {http://dx.doi.org/10.1063/1.4891884}
}

@article{Teubner_1997,
	author = {Teubner, U. and Altenbernd, D. and Gibbon, P. and Förster, E. and Mysyrowicz, A. and Audebert, P. and Geindre, J. P. and Gauthier, J. C. and Lichters, R. and Meyer-ter-Vehn, J.},
	title = {Observation of VUV radiation at wavelengths in the $\omega$_p$- and 2$\omega$_p$-wavelength range emitted from femtosecond laser-plasmas},
	journal = {Opt Comm},
	volume = {144},
	pages = {217},
	year = {1997},
	doi = {},
	url = {10.1016/S00304018(97)00468-9}
}

@article{Teubner_1999,
	author = {Teubner, U. and Gibbon, P. and Altenbernd, D. and Oberschmidt, D. and Förster, E. and Mysyrowicz, A. and Audebert, P. and Geindre, J. P. and Gauthier, J. C.},
	title = {Plasma frequency and harmonic emission from fs-laser plasmas},
	journal = {Laser and Particle Beams},
	volume = {17},
	pages = {613},
	year = {1999},
	doi = {},
	url = {https://doi.org/10.1017/S0263034699174056}
}

@article{Yugami,
	author = {Yugami, N. and Higashiguchi, T. and Gao, H. and Sakai, S. and Takahashi, K. and Ito, H. and Nishida, Y.},
	title = {Experimental Observation of Radiation from Cherenkov Wakes in a Magnetized Plasma},
	journal = {Phys. Rev. Lett.},
	volume = {89},
	pages = {065003},
	year = {2002},
	doi = {10.1103/PhysRevLett.89.065003},
	url = {http://dx.doi.org/10.1103/PhysRevLett.89.065003}
}

@article{Yoshii,
	author = {Yoshii, J. and Lai, C. H. and Katsouleas, T. and Joshi, C. and Mori, W. B.},
	title = {Radiation from Cerenkov Wakes in a Magnetized Plasma},
	journal = {Phys. Rev. Lett.},
	volume = {79},
	pages = {4194},
	year = {1997},
	doi = {10.1103/PhysRevLett.79.4194},
	url = {}
}

@article{Yip,
	author = {Yip, W. K.},
	title = {Theory of polarized type III and U burst emissions from the solar corona},
	journal = {Planetary and Space Science},
	volume = {18},
	pages = {867},
	year = {1970},
	doi = {},
	url = {https://doi.org/10.1016/0032-0633(70)90084-X}
}

@article{Hinkel,
	author = {Hinkel-Lipsker, D. E. and Fried, B. D. and Morales, G. J.},
	title = {Analytic expressions for mode conversion in a plasma with a linear density profile},
	journal = {Physics of Fluids B: Plasma Physics},
	volume = {4},
	pages = {559},
	year = {1992},
	doi = {},
	url = {https://doi.org/10.1063/1.860255}
}

@article{Malaspinal,
	author = {Malaspina, D. M. and Cairns, I. H. and Ergun, R. E.},
	title = {ANTENNA RADIATION NEAR THE LOCAL PLASMA FREQUENCY BY LANGMUIR WAVE EIGENMODES},
	journal = {The Astrophysical Journal},
	volume = {755},
	pages = {45},
	year = {2012},
	doi = {},
	url = {https://ui.adsabs.harvard.edu/abs/2012ApJ...755...45M/abstract}
}

@article{Edney,
	author = {Edney, S. D. and Robinson, P. A.},
	title = {Analytic treatment of electromagnetic emission near the plasma frequency via Langmuir wave decay},
	journal = {Phys. Plasmas},
	volume = {6},
	pages = {3799},
	year = {1999},
	doi = {},
	url = {https://doi.org/10.1063/1.873644}
}

@article{Cairns1987,
	author = {Cairns, I. H.},
	title = {Fundamental plasma emission involving ion sound waves},
	journal = {J. Plasma Physics},
	volume = {38},
	pages = {169},
	year = {1987},
	doi = {},
	url = {}
}

@article{Robinson,
	author = {Robinson, P. A.  and Willes, A. J. and Cairns, I. H.},
	title = {Fundamental plasma emission involving ion sound waves},
	journal = {Astrophys. J.},
	volume = {408},
	pages = {720},
	year = {1993},
	doi = {},
	url = {https://doi.org/10.1086/172632}
}

@article{Palastro,
	author = {Palastro, J. P. and Williams, E. A. and Hinkel, D. E. and Divol, L. and Strozzi,D. J. },
	title = {Kinetic dispersion of Langmuir waves. I. The Langmuir decay instability},
	journal = {Phys. Plasmas},
	volume = {16},
	pages = {092304},
	year = {2009},
	doi = {},
	url = {https://doi.org/10.1063/1.3234245}
}

@article{Marques2020,
	author = {Marquès, J.-R. and Briand, C. and Amiranoff, F. and Depierreux, S. and Grech, M. and Lancia, L. and Pérez, F. and Sgattoni, A. and Vinci, T. and Riconda, C.},
	title = {Laser-Plasma Interaction Experiment for Solar Burst Studies},
	journal = {Phys. Rev. Lett.},
	volume = {124},
	pages = {135001},
	year = {2020},
	doi = {},
	url = {https://doi.org/10.1103/PhysRevLett.124.135001}
}

@article{Perez,
	author = {Pérez, F. and Amiranoff, F. and Briand, C. and Depierreux, S. and Grech, M. and Lancia, L. and Loiseau, P. and Marquès, J.-R. and Riconda, C.and Vinci, T.},
	title = {Numerical study of Langmuir wave coalescence in laser-plasma interaction},
	journal = {Phys. Plasmas},
	volume = {28},
	pages = {043102},
	year = {2021},
	doi = {},
	url = {https://doi.org/10.1063/5.0037028}
}

@article{Albertazzi,
	author = {{Albertazzi}, B. and others},
	title = {Production of large volume, strongly magnetized laser-produced plasmas by use of pulsed external magnetic fields},
	journal = {Rev. Sci. Instrum.},
	volume = {84},
	pages = {043505},
	year = {2013},
	doi = {10.1063/1.4795551}
}

@article{Marques2025,
	author = {Marquès, J.-R. and Pérez, F. and Loiseau, P. and Lancia, L. and Briand, C. and Depierreux, S. and Grech, M. and Riconda, C.},
	title = {Experimental Evidence of Stimulated Raman Rescattering in Laser-Plasma Interaction},
	journal = {Phys. Rev. Lett.},
	volume = {135},
	pages = {105101},
	year = {2025},
	doi = {},
	url = {https://doi.org/10.1103/r7th-rq38}
}

@article{LaFontaine,
	author = {La Fontaine, B.  and Villeneuve, D. M. and Baldis, H. A.  and Drake, R. P. and Estabrook, K.},
	title = {Test of the Landau Cutoff of Stimulated Raman Scattering Spectra as an Electron-Temperature Diagnostic in Laser-Produced Plasmas},
	journal = {Phys. Rev. Lett.},
	volume = {68},
	pages = {484},
	year = {1992},
	doi = {},
	url = {https://doi.org/10.1103/PhysRevLett.68.484}
}

@article{TROLL,
	author = {Lefebvre, E. and Bernard, S. and Esnault, C. and Gauthier, P. and Grisollet, A. and Hoch, P. and Jacquet, L. and Kluth, G. and Laffite, S. and Liberatore, S. and Marmajou, I. and Masson-Laborde, P.-E. and Morice, O. and Willien, J.-L.},
	title = {Development and validation of the TROLL radiation-hydrodynamics code for 3D hohlraum calculations},
	doi = {10.1088/1741-4326/aacc9c},
	journal = {Nucl. Fusion},
	volume = {59},
	pages = {032010},
	year = {2019},
	url = {https://doi.org/10.1088/1741-4326/aacc9c}
}

@article{Smilei,
	author = {Derouillat, J. and Beck, A. and Pérez, F. and Vinci, T. and Chiaramello, M. and A.Grassi, A. and Flé, M. and Bouchard, G. and Plotnikov, I. and Aunai, N. and Dargent, J. and Riconda, C. and Grech, M.},
	title = {Smilei: A Collaborative, Open-Source, Multi-Purpose Particle-in-Cell Code for Plasma Simulation},
	doi = {10.1016/j.cpc.2017.09.024},
	journal = {Comput. Phys. Commun.},
	volume = {222},
	pages = {351},
	year = {2018},
	url = {https://doi.org/10.1016/j.cpc.2017.09.024}
}

@misc{SM,
	author = {},
	title = {See the supplementary material at ? for more experimental results and Particle-In-Cell and hydrodynamic simulations},
	howpublished = "\url{URL_will_be_inserted_by_publisher}",
	year = {},
	note = " "
}

\section*{End Matter}

\textbf{Experimental setup}

Light emitted from the plasma center at 25$^{\circ}$, 40$^{\circ}$, 75$^{\circ}$, 110$^{\circ}$ and 170$^{\circ}$ is resolved in time and in wavelength in the 700-900 nm domain using spectrometers coupled to streak cameras. A multiplexing system is used to record the light collected at 25$^\circ$ and at 40$^\circ$ on a single streak camera. A similar system combines the signals collected at 75$^\circ$ and 110$^\circ$. The volume probed is $L_x\times L_y\times L_z\sim 300\times 90\times90~\mu$m$^3$, except for the 170$^{\circ}$ diagnostic for which it is $\sim 500\times300\times300~\mu$m$^3$. The spectral and temporal resolutions of those diagnostics are $\sim 5$ nm and $\sim 0.1$ ns. The interaction region is imaged onto a GOI camera placed at 85$^{\circ}$. The image is time-integrated on a 6 ns gate centered on the laser pulse. All these diagnostics cover the spectral range $700-900$ nm. The light collected at 75$^\circ$ and 110$^\circ$ is also recorded on an InGaAs camera to image the interaction region in the 1100-1700 nm spectral range. To achieve this, a 5 mm thick silicon plate is placed in front of the spectrometer. Wavelengths below 1100 nm are reflected towards the spectrometer, and wavelengths above 1100 nm are transmitted to the camera. The camera integrates over 10 µs.

\begin{figure}[h!]
	\begin{center}
		\includegraphics[width=\columnwidth]{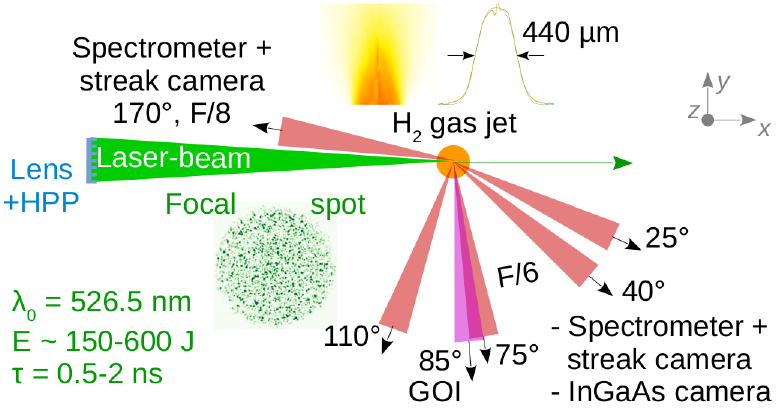}
		\caption{\small{\textit{Scheme of the experimental setup showing the angular positions of the diagnostics, the laser's focal spot image, and the gas jet profile.}}}
		\label{Fig_Setup}
	\end{center}
\end{figure}

\textbf{Visible-NIR sideview}

\begin{figure}[h!]
	\begin{center}
		\includegraphics[width=0.6\columnwidth]{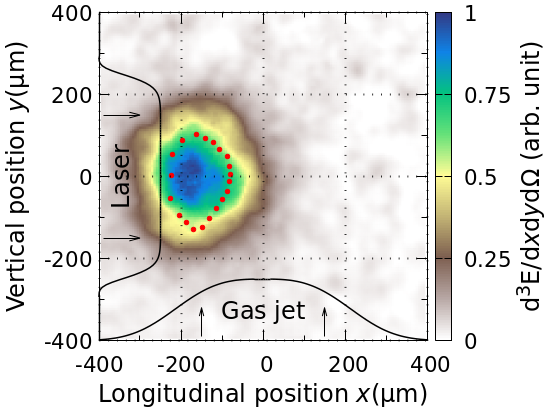}
		\caption{\small{\textit{2D-images of the interaction region obtained with the GOI camera, from the scattered light emitted at 85$^\circ$ (side view) of the laser axis. It shows the same interaction region than Fig. \ref{Fig_InGaAs_image}, but in the 700-900 nm wavelength domain associated with the Raman and $2\omega_p$ spectra shown in Fig. \ref{Fig_Spectra}. The laser propagates from left to right. The black curves show the transverse profile of the gas jet and of the laser focal spot (averaged on the speckles, see Fig. \ref{Fig_Setup}). The red dots indicate the area associated with the $\omega_p$ emission, measured with the InGaAs camera at 75$^{\circ}$ and 110$^{\circ}$, shown in Fig. \ref{Fig_InGaAs_image}.
		}}}
		\label{Fig_Sideview}
	\end{center}
\end{figure}

Figure \ref{Fig_Sideview} shows the EM emission from the GOI diagnostic placed at 85$^\circ$ (see Fig. \ref{Fig_Setup}). The image is time integrated on a 6 ns gate centered on the laser pulse and covers the 700-900 nm wavelength domain, corresponding to the SRS and $2\omega_p$ emissions observed on the spectrally and time resolved diagnostics presented in Fig. \ref{Fig_Spectra}. The spatial extent of these emissions is greater than that at $\omega_p$ measured with the InGaAs camera at 75$^{\circ}$ and 110$^{\circ}$ (see Fig. \ref{Fig_InGaAs_image}), indicated by the red dots.

\begin{figure}
	\begin{center}
		\includegraphics[width=\columnwidth]{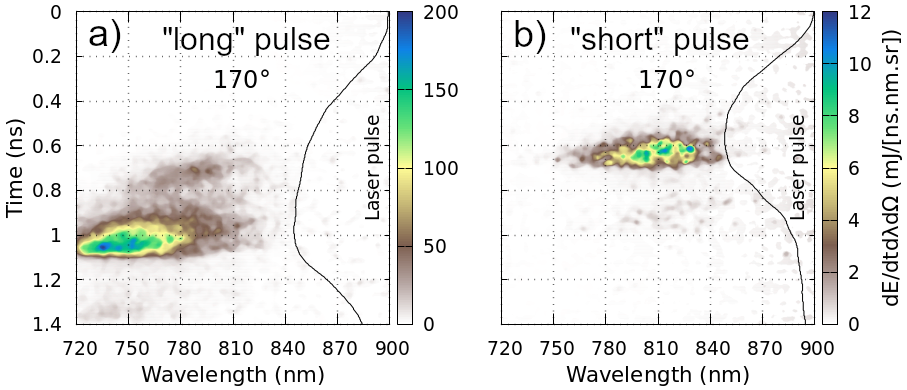}
		\caption{\small{\textit{Time-resolved spectra measured at 170$^\circ$ from the laser axis, for two same shots than in Fig. \ref{Fig_Spectra}, one with a "long" laser pulse, a), and one with a "short" pulse, b). The black curve on the right of each graph shows the temporal profile of the laser. Each spectrum is corrected from the spectral response of the diagnostic.}}}
		\label{Fig_Spectra_BRS}
	\end{center}
\end{figure}

\textbf{Time-resolved spectra measured at 170$^\circ$}
Figure \ref{Fig_Spectra_BRS} shows the time-resolved frequency spectra of the Raman scattered light emitted at 170$^\circ$ for two laser pulse durations, one "long" and one "short". In the case of the "short" pulse, Fig. \ref{Fig_Spectra}-b and \ref{Fig_Spectra}-d, the SRS emission stops at a wavelength near 780 nm. This indicates that the excitation of Langmuir waves is restricted to plasma densities between 0.25 and 0.1 $n_c$

\end{document}